\documentclass[aip,amsmath,amssymb,reprint]{revtex4-1}
\usepackage[utf8]{inputenc}
\usepackage{dcolumn}
\usepackage{bm}
\usepackage{graphicx}
\usepackage[utf8]{inputenc}
\usepackage[T1]{fontenc}
\usepackage{mathptmx}
\usepackage{etoolbox}

\usepackage{color}
\usepackage{orcidlink}
\usepackage{lineno}
\usepackage{listings} 
\definecolor{lightergray}{gray}{0.95}
\colorlet{darkorange}{orange!80!black}
\lstdefinelanguage{bash_deepqmc}{
    language=bash,
    morekeywords={deepqmc},
}

\usepackage[normalem]{ulem}

\usepackage{physics}

\makeatletter
\def\@email#1#2{%
 \endgroup
 \patchcmd{\titleblock@produce}
  {\frontmatter@RRAPformat}
  {\frontmatter@RRAPformat{\produce@RRAP{*#1\href{mailto:#2}{#2}}}\frontmatter@RRAPformat}
  {}{}
}%
\makeatother
\begin{document}


\title{DeepQMC: an open-source software suite for variational optimization of deep-learning molecular wave functions.}
\author{Z. Schätzle \orcidlink{0000-0002-5345-6592}}
\thanks{Z. Schätzle and P. B. Szabó contributed equally to this work.}
\affiliation{FU Berlin, Department of Mathematics and Computer Science,
Arnimallee 6, 14195 Berlin, Germany}
\author{P. B. Szabó \orcidlink{0000-0003-1824-8322}}
\thanks{Z. Schätzle and P. B. Szabó contributed equally to this work.}
\affiliation{FU Berlin, Department of Mathematics and Computer Science,
Arnimallee 6, 14195 Berlin, Germany}
\author{M. Mezera \orcidlink{0009-0003-0047-488X}}
\affiliation{FU Berlin, Department of Mathematics and Computer Science,
Arnimallee 6, 14195 Berlin, Germany}
\author{J. Hermann \orcidlink{0000-0002-2779-0749}}
\affiliation{Microsoft Research AI4Science, Karl-Liebknecht Str. 32, 10178 Berlin, Germany}
\author{F. Noé \orcidlink{0000-0003-4169-9324}}
\affiliation{FU Berlin, Department of Mathematics and Computer Science,
Arnimallee 6, 14195 Berlin, Germany}
\affiliation{Microsoft Research AI4Science, Karl-Liebknecht Str. 32, 10178 Berlin, Germany}
\affiliation{FU Berlin, Department of Physics, Arnimallee 14, 14195 Berlin, Germany}
\affiliation{Rice University, Department of Chemistry, Houston, TX 77005, USA}
\date{\today}
\begin{abstract}
    Computing accurate yet efficient approximations to the solutions of the electronic Schr\"odinger equation has been a paramount challenge of computational chemistry for decades.
    Quantum Monte Carlo methods are a promising avenue of development as their core algorithm exhibits a number of favorable properties: it is highly parallel, and scales favorably with the considered system size, with an accuracy that is limited only by the choice of the wave function ansatz.
    The recently introduced machine-learned parametrizations of quantum Monte Carlo ansatzes rely on the efficiency of neural networks as universal function approximators to achieve state of the art accuracy on a variety of molecular systems.
    With interest in the field growing rapidly, there is a clear need for easy to use, modular, and extendable software libraries facilitating the development and adoption of this new class of methods.
    In this contribution, the \textsc{DeepQMC} program package is introduced, in an attempt to provide a common framework for future investigations by unifying many of the currently available deep-learning quantum Monte Carlo architectures.
    Furthermore, the manuscript provides a brief introduction to the methodology of variational quantum Monte Carlo in real space, highlights some technical challenges of optimizing neural network wave functions, and presents example black-box applications of the program package. 
    We thereby intend to make this novel field accessible to a broader class of practitioners both from the quantum chemistry as well as the machine learning communities.
\end{abstract}

\maketitle

\section{Introduction}
Recently, the application of machine learning to a wide range of problems from the natural sciences has proven to be highly successful.
Computational chemistry is a field of particular activity: machine learning force fields model complicated quantum mechanical effects at the resolution of atoms, while machine learned functionals elevate density functional theory to unprecedented accuracy ~\cite{noe_machine_2020, kulik_roadmap_2022, kirkpatrick_pushing_2021}.
These approaches utilize supervised training to learn from accurate quantum mechanical reference calculations, and make predictions for unseen configurations.
While this results in fast yet accurate approximations to the quantum many-body problem, it inherently depends on high quality training data, which represents a major bottleneck of these methods.

An orthogonal way to incorporate machine learning into computational chemistry is its application to improve ab-initio calculations. Notably, over the course of the last years a new family of deep-learning quantum Monte Carlo (deep QMC) methods has developed, incorporating advancements from the field of machine learning ~\cite{hermann_ab-initio_2022}. 
Common to the deep QMC methods is the utilization of neural networks to parametrize highly expressive ansatzes, efficiently approximating the solutions of the time-independent electronic Schr\"odinger equation, thereby providing a complete description of the system's electronic properties.
Originating from spin lattices~\cite{carleo_solving_2017}, deep-learning ansatzes were soon applied to molecules in real-space~\cite{han_solving_2019}. With the development of PauliNet~\cite{hermann_deep_2020} and FermiNet~\cite{pfau_ab_2020}, the accuracy of neural-network wave functions became the state of the art within variational Monte Carlo. Subsequent works have further increased the accuracy of these ansatzes~\cite{schatzle_convergence_2021,lin_explicitly_2023, gerard_gold-standard_2022, von_glehn_self-attention_2022}, extended them to the simulation of excited states~\cite{entwistle_electronic_2023} as well as periodic systems~\cite{pescia_neural-network_2022,wilson_neural_2023}, combined them with pseudo-potentials~\cite{li_fermionic_2022}, used them in the calculation of interatomic forces~\cite{qian_interatomic_2022}, utilized them in diffusion Monte Carlo simulations~\cite{wilson_simulations_2021, ren_towards_2023}, and extended them to share parameters across multiple molecular geometries~\cite{scherbela_solving_2021, gao_ab-initio_2022, gao_sampling-free_2023} or distinct molecules~\cite{scherbela_towards_2023, gao_generalizing_2023}.

Although the method of optimizing deep-learning wave function ansatzes using variational quantum Monte Carlo was developed only a few years ago, it already competes with some of the most accurate traditional quantum chemistry methods on molecules with up to $\sim$100 electrons.
Exhibiting competitive scaling with the number of electrons, it has the potential to be extended to larger systems in the near future. Achieving this will no doubt require further method development as well as efficient implementations of the core algorithms, creating the need for open source libraries that facilitate experimentation and contribution from the community.

Accompanying the above summarized research, various software libraries for variational optimization of deep-learning wave functions have been released~\cite{netket3_2022, deepqmc_2023, ferminet_2020, deeperwin_2022}. While \textsc{NetKet}~\cite{netket3_2022} provides a general implementation of variational optimization of machine learning wave functions mainly for lattice systems with recent extensions to continuous space, the research for molecular machine learning wave functions was carried out across various repositories and is lacking a unified framework.
The presented \textsc{DeepQMC} program package aims to provide a unified implementation of the developments in the field of deep-learning molecular wave functions. It intends to be easy to use out of the box, while allowing full control and flexibility over the ansatz and variational training for advanced users. The library is designed to be modular, facilitating the rapid development and testing of individual components, and easing the implementation of new features.
It makes use of the composable function transformations and just-in-time compilation provided by the JAX library~\cite{jax2018github} to express performant GPU accelerated algorithms using concise Python~\cite{van_rossum_python3} syntax.
Neural network models are encapsulated in higher-level objects, using the \texttt{haiku} deep-learning library~\cite{haiku2020github}.
The project is open-source and distributed online under the MIT license~\cite{deepqmc_2023}.

\section{Theory}
\subsection{The electronic structure problem}
In computational chemistry, molecular systems are often described by the non-relativistic molecular Hamiltonian using the Born--Oppenheimer approximation:
\begin{equation}
\label{eq:molecular_hamiltonian}
    \hat{H} = \sum_{i=1}^N\Big(-\frac12\Delta_{\mathbf r_i} - \sum_{I=1}^M\tfrac{Z_I}{|\mathbf{r}_i-\mathbf{R}_I|} + \sum_{j=1}^{i-1} \tfrac{1}{|\mathbf{r}_i - \mathbf{r}_j|}\Big ) \, ,
\end{equation}
where ${\bf r}_i$ denotes the position of the $i$th electron, while $Z_I$ and ${\bf R}_I$ are the charge and position of the $I$th nucleus, respectively.
To determine the electronic structure of these molecular systems, one must solve the associated time independent Schr\"odinger equation
\begin{equation}
\label{eq:schroedinger_equation}
    \hat{H}\psi(\mathbf x_1, ..., \mathbf x_N) = E\psi (\mathbf x_1, ..., \mathbf x_N) \, ,
\end{equation}
where $\mathbf x_i=(\mathbf r_i, \sigma_i)$ comprise the positions of the electrons and their spin.
A solution is an eigenfunction of the Hamiltonian, the electronic wave function $\psi$, and its corresponding energy eigenvalue $E$.
With the exact wave function at hand, any observable electronic property of the system can in principle be computed, as the wave function gives a complete description of the system's electronic state.
Since electrons have half-integer spin, their wave functions must be antisymmetric with respect to electron exchanges
\begin{equation}
\label{eq:antisymmetry}
\begin{split}      
    \psi(\ldots,\mathbf x_i,\ldots,\mathbf x_j,\ldots)=  -\psi(\ldots,\mathbf x_j,\ldots,\mathbf x_i,\ldots) \; .
\end{split}
\end{equation}
While general wave functions are complex-valued, the solutions of the time independent Schr\"odinger equation can be chosen as real without loss of generality, due to the hermiticity of the molecular Hamiltonian.
Therefore in all of the following discussions, as well as in the \textsc{DeepQMC} library, only real valued wave functions are considered.

\subsection{Variational optimization}\label{sec:variational}
Even with the aforementioned approximations, the electronic Schrödinger equation involving the molecular Hamiltonian can only be solved analytically for hydrogenic atoms -- the special case of the two-body problem. This makes computational quantum chemistry a mainly numerical field, where different methods yield approximate solutions at varying trade-offs of accuracy and computational cost. The class of variational quantum chemistry methods phrases the solution of the Schrödinger equation as a minimization problem. The ground state of the Hamiltonian is approximated by optimizing the parameters $\boldsymbol \theta$ of a trial wave function (ansatz) $\psi_{\boldsymbol \theta}$, to minimize the expectation value of the Hamiltonian 
\begin{equation}
    \boldsymbol\theta' = \underset{\boldsymbol\theta}{\text{argmin}} \langle \hat{H}\rangle_{\psi_{\boldsymbol\theta}}.
\end{equation}
This objective is rooted in the variational principle of quantum mechanics, which states that the ground state energy of the Hamiltonian is a lower bound for the energy expectation value of any wave function from the associated antisymmetric Hilbert space $H^-$
\begin{equation}
\label{eq:variational_prnciple}
    E_{0} \leq \underset{\psi}{\text{min}} \langle \hat{H}\rangle_\psi  \;\;\; \psi \in H^-.
\end{equation}
The variational methods can be categorized based on the means of calculating the expectation value $\langle\cdot\rangle$, and choice of ansatz $\psi_{\boldsymbol \theta}$.

The \textsc{DeepQMC} program package implements VMC in real space (first quantization) with neural network wave functions. In VMC, expectation values are estimated through a stochastic sampling of electron configurations 
\begin{equation}
\label{eq:expectation_values}
\begin{split}      
    \langle \hat{H}\rangle_{\psi_{\boldsymbol{\theta}}} &=
    \frac{\langle \psi_{\boldsymbol{\theta}} |\hat{H}|\psi_{\boldsymbol{\theta}} \rangle}{\langle \psi_{\boldsymbol{\theta}} |\psi_{\boldsymbol{\theta}} \rangle}\\
    &= \frac{\int d\mathbf r_1, ..., d\mathbf r_N \psi_{\boldsymbol{\theta}}^*(\mathbf r_1, ..., \mathbf r_N)\hat{H}\psi_{\boldsymbol{\theta}}(\mathbf r_1, ..., \mathbf r_N)}{\int d\mathbf r_1, ..., d\mathbf r_N |\psi_{\boldsymbol{\theta}}(\mathbf r_1, ..., \mathbf r_N)|^2}\\
    &= \frac{\int d\mathbf r_1, ..., d\mathbf r_N |\psi_{\boldsymbol{\theta}}(\mathbf r_1, ..., \mathbf r_N)|^2E_{\text{loc}}[\psi_{\boldsymbol\theta}](\mathbf r)}{\int d\mathbf r_1, ..., d\mathbf r_N |\psi_{\boldsymbol{\theta}}(\mathbf r_1, ..., \mathbf r_N)|^2}\\
    &= \mathbb{E}_{\mathbf r \sim |\psi_{\boldsymbol\theta}|^{2}}\big[E_{\text{loc}}[\psi_{\boldsymbol\theta}](\mathbf r)\big]\\
    &\approx \frac{1}{n} \sum^n_{{\bf r} \sim |\psi_{\boldsymbol \theta}|^2}E_\text{loc}[\psi_{\boldsymbol \theta}]({\bf r})\,.
\end{split}
\end{equation}
Because the molecular Hamiltonian does not depend on the spin, it is possible to compute the energy using the spatial wave function $\psi(\mathbf r_1, ..., \mathbf r_N)$, where fixed spins are assigned to the electrons and spin-up and spin-down electrons are treated as distinguishable~\cite{foulkes_quantum_2001}. The convention is to sort the electrons by spin and consider the first $N^\uparrow$ electrons to have spin-up and the remaining $N^\downarrow=N-N^\uparrow$ electrons to possess spin-down.

In practice, a VMC simulation then consists of choosing an ansatz (see Section\,\ref{sec:design_space}), and optimizing it in an alternating scheme of sampling and parameter updates.
The expectation value in \eqref{eq:expectation_values} is approximated by sampling the probability density given by the square of the wave function (see Section\,\ref{sec:sampling}), followed by a parameter update using the gradient of the expectation value (see Section\,\ref{sec:training}).

\subsection{Neural network wave functions}
Being exact in principle, the choice of the wave function ansatz is crucial for the efficiency of a VMC simulation.
Recently, neural network parametrizations of real-space molecular wave functions were introduced by PauliNet~\cite{hermann_deep_2020}, and FermiNet~\cite{pfau_ab_2020}.
They both rely on generalized Slater determinants, that augment the single particle orbitals of conventional Slater determinants with many-body correlation,
\begin{equation}\label{eq:dl_ansatz}
    \psi_{\boldsymbol\theta}(\mathbf r_1, ..., \mathbf r_N) =
  \textstyle{\sum}_p c_p
  \det[\mathbf A^p(\mathbf r)] \, ,
\end{equation}
\begin{equation}\label{eq:mb_orbital}
    A^p_{ik}= \phi^p_k(\mathbf r_i, \{\mathbf r^\uparrow\},\{\mathbf r^\downarrow\}) \times \varphi^p_k(\mathbf r_i) \, .
\end{equation}
Here, $\phi^p_k$ are many-body orbitals, and $\varphi^p_k$ are single particle envelopes that ensure the correct asymptotic behavior of the wave function with increasing distance from the nuclei. 
The set notation $\{\cdot\}$ is to be understood as a permutation invariant dependence on the spin-up electrons $\mathbf r^\uparrow$, and spin-down electrons $\mathbf r^\downarrow$, respectively.
The ansatz may be a linear combination of multiple generalized Slater determinants, which are distinguished with the $p$ index. The form of $\phi^p_k$ in \eqref{eq:mb_orbital} is closely related to the backflow transformation~\cite{feynman_energy_1956}, which introduces quasi-particles to obtain many-body orbital functions.
The key observation motivating this augmentation is that the antisymmetry of Slater determinants constructed from many-body orbitals is preserved as long as the orbital functions are equivariant with the exchange of electrons,
\begin{equation}\label{eq:perm_cond}
    P^\parallel_{ij} \phi_k(\mathbf r_i, \{\mathbf r^\uparrow\},\{\mathbf r^\downarrow\}) = \phi_k(\mathbf r_j, \{\mathbf r^\uparrow\},\{\mathbf r^\downarrow\}) \, ,
\end{equation}
where $P^\parallel_{ij}$ is the operator exchanging same-spin electrons $i$ and $j$.

Most of the currently used deep-learning molecular wave functions~\cite{hermann_deep_2020, pfau_ab_2020, gerard_gold-standard_2022} share the functional form of \eqref{eq:dl_ansatz}, and \eqref{eq:mb_orbital}, and differ only in the parametrization of the many-body orbitals $\phi^p_k$ and single-particle envelopes $\varphi^p_k$.
\textsc{DeepQMC} aims to provide a general framework for variational optimization of deep-learning molecular wave functions, facilitating the investigation of the design space spanned by the PauliNet, FermiNet, and DeepErwin neural network ansatzes. 

\subsection{Pseudopotentials}\label{sec:pseudopotentials}
Despite the favorable asymptotic scaling of VMC with the number of electrons, systems containing heavy nuclei remain challenging due to a variety of reasons.
The high energy of electrons near these nuclei complicates simulations by spoiling the optimization and reducing the effectiveness of MCMC sampling.
Furthermore, the kinetic energy of these electrons reaches the relativistic regime, requiring the treatment of relativistic effects that are not accounted for in the standard non-relativistic molecular Hamiltonian of \eqref{eq:molecular_hamiltonian}.
On the other hand, while the core regions of heavily charged nuclei contribute dominantly to the total energy, they are typically unchanged during chemical processes and thus have little effect on the chemically relevant relative energies.
Therefore most quantum chemistry methods targeted at computing relative energies reduce the above outlined difficulties, by treating the outer (valence) electrons separately from the inner (core) electrons.

The approach most suited for implementation in the context of variational optimization of deep-learning wave functions is the use of pseudopotentials, which has been previously demonstrated by Li et al.~\cite{li_fermionic_2022}.
In this method, the core electrons are excluded from the explicit calculation and replaced by additional terms in the Hamiltonian, to simulate their influence on the remaining $N_\text{v}$ valence electrons. 
The modified Hamiltonian reads as
\begin{equation}
\label{eq:modified_molecular_hamiltonian}
    \hat{H} = \sum_{i=1}^{N_\text{v}}\Big(-\frac12\Delta_{\mathbf r_i} + \sum_{j=1}^{i-1} \tfrac{1}{|{\mathbf r}_i - {\mathbf r}_j|} \Big ) + \hat V_{\text{PP}}\,.
\end{equation}
The $\hat V_\text{PP}$ pseudopotential term is in turn decomposed to local and non-local parts
\begin{equation}
    \label{eq:pseudopotential}
    \hat{V}_{\text{PP}} = \sum\limits_{I=1}^{M}  \sum\limits_{i=1}^{N_\text{v}}\left( V^I(r_{iI}) + \sum\limits_{l=0}^{l_{\text{max}}} W_l^I (r_{iI}) \hat{P}_l^{iI}\right),
\end{equation}
where $r_{iI}=\abs{\mathbf{r}_i - \mathbf{R}_I}$, $V^I$ and $W^I_l$ are sets of scalar functions describing the local and non-local pseudopotential contributions, while $\hat{P}_l^{iI} = \sum_{m=-l}^{l}\ket{lm}_{iI}\bra{lm}_{iI}$ is a projection operator of the $i$-th electron on spherical harmonics centered on the $I$-th nucleus.
To evaluate the contribution of the non-local part of the pseudopotential (second term of \eqref{eq:pseudopotential}) one considers integrals of the form

\begin{equation}
\begin{split}
    \frac{\bra{\mathbf r} W^I_l \hat{P}^{iI}_{l} \ket{\psi}}{\braket{\mathbf r}{\psi}} =
     W^I_l&(r_{iI})\!\!\! \sum\limits_{m=-l}^l \!\! Y_{lm}({\scriptstyle \boldsymbol \Omega_{iI}})\\
    \times\!\!\int\! Y_{lm} ({\scriptstyle \boldsymbol \Omega'_{iI}})^{*}&\frac{\psi({\scriptstyle \mathbf r_1, ..., (r_{iI}, \boldsymbol \Omega'_{iI}), ..., \mathbf r_{N_\mathrm{v}}})}{\psi({\scriptstyle \mathbf r_1, ..., (r_{iI}, \boldsymbol \Omega_{iI}), ..., \mathbf r_{N_\mathrm{v}}})}\dd {\scriptstyle \boldsymbol \Omega'_{iI}}
\end{split}
\label{eq:pp_integral}
\end{equation}
where $Y_{lm}$ is a spherical harmonic and $({r_{iI}, \boldsymbol \Omega_{iI}})$ denotes the position vector of the $i$-th electron $\mathbf{r}_i$, expressed in spherical coordinates centered on nucleus $I$. 
Following the first implementation of pseudopotentials for deep-learning molecular wave functions by Li et al.~\cite{li_fermionic_2022}, the above integral is approximated using an icosahedral quadrature of 12-points.

The scalar functions $V^I$ and $W^I_l$ are typically pre-computed by expanding them in a Gaussian basis, and fitting the expansion parameters directly to a database of reference energies.
The \textsc{DeepQMC} program package currently includes the widely used BFD~\cite{burkatzki_energy-consistent_2007}, and the most recent ccECP~\cite{benett_new_2017} pseudopotentials, with an application of the latter presented in Section\,\ref{sec:transition_metal_oxides}.

\section{Wave function design space}\label{sec:design_space}
\textsc{DeepQMC} implements a variety of options to obtain the equivariant many-body orbitals $\phi^p_k$ and the accompanying envelopes $\varphi^p_k$, covering PauliNet, FermiNet, DeepErwin and their derivatives. In the following, the main architectural concepts of these wave function ansatzes will be described in more detail. For ease of use \textsc{DeepQMC} provides predefined configuration files to obtain the above mentioned ansatzes, while allowing their interpolation through a manual choice of hyperparameters.

\subsection{Graph Neural Networks}\label{sec:gnn}
\begin{figure*}[t]
    \centering
    \includegraphics[width=\textwidth]{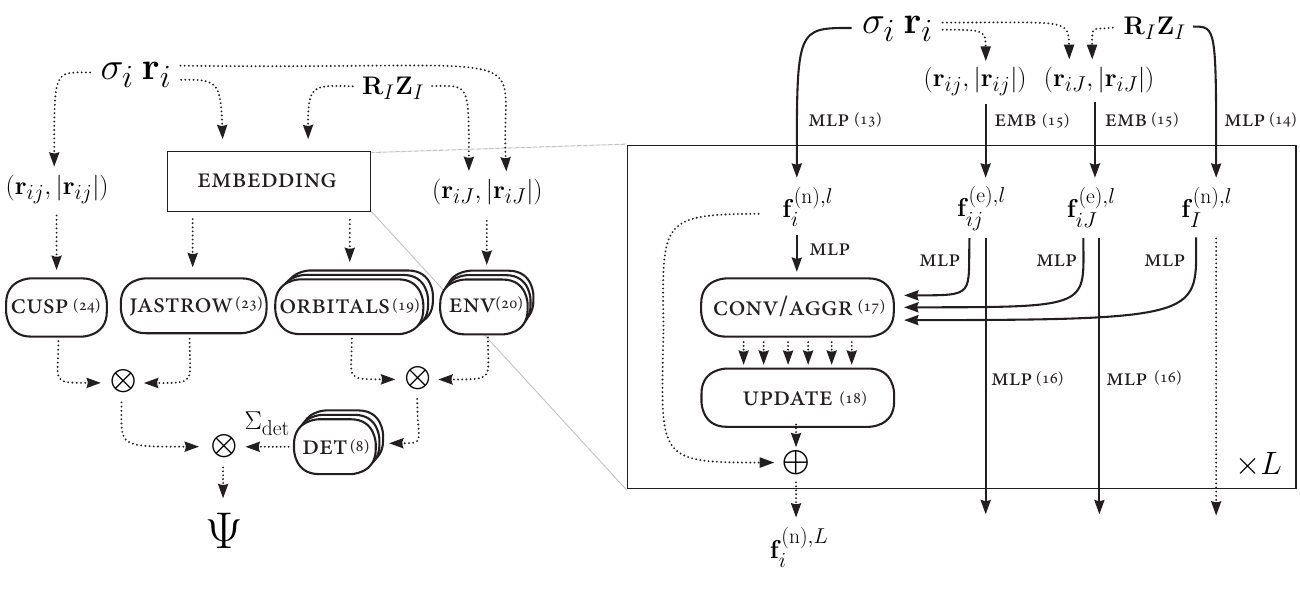}
    \caption{\textbf{Sketch of a general neural network wave function ansatz and its graph neural network architecture.} This sketch comprises the implemented design space for the neural network wave function (left) and the GNN architecture (right). Solid lines can carry MLPs and dotted lines correspond to forwarding without further change. Numbers in parentheses refer to the corresponding equations of the main text. The choice among the drawn connections, the depths of the associated MLPs as well as the aggregation and update rules distinguish the previously published works PauliNet, FermiNet, and DeepErwin.}
    \label{fig:design_space}
\end{figure*}
Central to the neural network wave function ansatzes is the computation of equivariant many-body embedding vectors for the electrons, which are used downstream to obtain the entries of the generalized Slater determinant.
Many strategies of obtaining these embeddings can be unified in the framework of graph neural networks (GNNs). 
GNNs are well suited to model functions on graphs that exhibit certain permutational symmetries, and can be adapted to describe electrons of molecules in real-space.

An electronic configuration of a molecule can be encoded as a graph, where the nodes are electrons and nuclei, and the connecting edges carry pairwise features, e.g. difference vectors.
GNNs are functions of these graphs, yielding high-dimensional latent space embeddings of the nodes.
The electronic node embeddings are initialized with single-electron features and iteratively updated to incorporate many-body information of the electronic environment.
Using graph convolutions, the updates are invariant under the exchange of electrons in the environment, and the conditions of \eqref{eq:perm_cond} are fulfilled.

The most relevant aspects of the GNN architecture implemented in \textsc{DeepQMC} are sketched on the right pane of Figure \ref{fig:design_space}, and are discussed in detail in the following.
Electron positions (spins) are denoted with $\bf r$ ($\sigma$), while $\bf R$ and $Z$ indicate nuclei positions and charges.
Node and edge quantities are denoted with the superscripts ($\text{n}$) and ($\text{e}$), respectively.
Furthermore, $l$ indexes the GNN interaction layers, $\boldsymbol \theta$ denotes functions parametrized by MLPs, and $t$ runs over the different edge types (those between electrons of identical or opposite spins, or between electrons and nuclei), node types (electron or nuclei nodes) or message types.
Lastly, vertical brackets indicate the different options implemented in \textsc{DeepQMC} for the computation of various quantities.

\subsubsection*{The graph representation:}
A graph is a natural way to denote the electronic configuration of a molecule in real-space. The nodes of the graph represent particles (electrons and nuclei), carrying information such as spin or nuclear charge. The edges support the difference vectors between the particles, resulting in a representation invariant under global translation. Note that using internal coordinates that are invariant under global rotation may not be sufficient to represent all wave functions (simple counterexamples are atomic wave functions with P symmetry), and can only be employed with a careful treatment of the spatial symmetries. 

To implement a variety of wave function ansatzes, \textsc{DeepQMC} provides configuration options to define the specifics of the graph construction outlined above.
Most importantly, the nodes corresponding to nuclei, and their respective nuclei-electron edges can optionally be excluded from the graph.
In this case, electron-nuclei information can still be introduced to the GNN, by initializing the electron embeddings using a concatenation of the difference vectors between the positions of the given electron and all nuclei (see the second case of \eqref{eq:el_init}).

\subsubsection*{Node features:}
The output of \textsc{DeepQMC} GNNs are electron node embeddings $\mathbf f_i^\text{(n)}$ which are subsequently used to generate the many-body orbitals that constitute the entries of the generalized Slater determinants and an optional trainable Jastrow factor.
To enforce the equivariance of these quantities with respect to the exchange of electrons, the initialization of the electron embeddings has to be chosen appropriately. In \textsc{DeepQMC} one can either use identical embeddings for all electrons of the same spin (invariant under permutation of same-spin electrons), or a concatenation of the electron nuclei difference vectors (equivariant under electron permutations)
 \begin{equation}
    \mathbf f_i^{(\text{n}),0,\text{el}} = 
    \begin{cases} 
     ~\mathcal{E}^{(\text{n}),\text{el}}_{\boldsymbol\theta}(\sigma_i) \\
     ~\mathcal{E}^{(\text{n}),\text{el}}_{\boldsymbol\theta}(\mathbf{r}_i - \mathbf{R_1} , ..., \mathbf{r}_i - \mathbf{R_N} )\,,
\end{cases}
\label{eq:el_init}
 \end{equation}
 where $\mathcal{E}^{(\text{n}),\text{el}}_{\boldsymbol\theta}$ are parameterized node embedding functions implemented through MLPs.

If the GNN is chosen to explicitly consider electron-nuclei interactions, the embeddings associated with the nuclear nodes have to be initialized besides the electronic embeddings. \textsc{DeepQMC} implements fixed nuclear node embeddings $\mathbf f_I^{(\text{n})}$, that either distinguish all nuclei or depend on the respective atom type:
 \begin{equation}
 \mathbf f_I^{(\text{n}),0,\text{nuc}} =
 \begin{cases} 
     ~\mathcal{E}^{(\text{n}),\text{nuc}}_{\boldsymbol\theta}(I) \\
     ~\mathcal{E}^{(\text{n}),\text{nuc}}_{\boldsymbol\theta}(Z_I)\,.
\end{cases}
 \end{equation}

 \subsubsection*{Edge features :}
The edges of the graph hold the pairwise differences of node positions ($\mathbf r_{ij}$) and their embeddings are consequently initialized as
 \begin{equation}
    \mathbf f_{ij}^{(\text{e}),0,{t^\text{(e)}}} = \mathcal{E}^{(\text{e}),{t^\text{(e)}}}(\mathbf{r}_{ij}) \, ,
 \end{equation}
 where $\mathcal{E}^{(\text{e}),{t^\text{(e)}}}$ is an edge type dependent featurization based on the pairwise differences. This may correspond to directly feeding the difference vectors, using the pairwise distances, expanding in a basis of Gaussians or working with rescaled difference vectors amongst other options.
 In later interaction layers, the original edge embeddings are either reused or iteratively updated,
\begin{equation}
    \mathbf f^{(\text{e}),l,{t^\text{(e)}}}_{ij} =
    \begin{cases}
        ~\mathbf f^{(\text{e}),0,{t^\text{(e)}}}_{ij} \\
        ~u_{\boldsymbol\theta}^{l,{t^\text{(e)}}}(\mathbf f^{(\text{e}),l-1,{t^\text{(e)}}}_{ij}) \, ,
    \end{cases}
\end{equation}
with the latter option making use of a parametrized update function $u_{\boldsymbol \theta}^{l, {t^\text{(e)}}}$, thus increasing the effective depth of the architecture at the cost of additional MLPs. The parameters of the update function $u_{\boldsymbol \theta}^{l, {t^\text{(e)}}}$ may be shared across different edge types.

\subsubsection*{Message generation:}
The electron embeddings are updated in each interaction layer by aggregating messages passed along the graph edges.
These messages are constructed via an elementwise product between functions of the sending node and edge embeddings (graph convolution),
\begin{equation}\label{eq:graphconvolution}
    \mathbf m_i^{l,{t^\text{(m)}}} = \sum_{j \in {t^\text{(n)}}} w_{\boldsymbol\theta}^{l,{t^\text{(e)}}}(\mathbf f^{(\text{e}),l,{t^\text{(e)}}}_{ij}) \cdot h_{\boldsymbol\theta}^{l,{t^\text{(n)}}}(\mathbf f^{(\text{n}),l,{t^\text{(n)}}}_{j}) \, .
\end{equation}
The superscript ${t^\text{(n)}}$ on the node features specifies the subset of sending nodes and the superscript ${t^\text{(e)}}$ on the edge features depends on the type of the sending and the receiving nodes respectively. The choice of how to distinguish electron-electron messages based on their spin (relative spin of sending and receiving electron, spin of sending electron, or no distinction between messages from spin-up and spin-down electrons) is another hyperparameter of the GNN.
Optionally, the above sum over the edges can be normalized by dividing it with the number of edges.
Note that messages depending only on node (edge) information can be obtained by setting the function $w^{l,{t^\text{(e)}}}_{\boldsymbol \theta}$ ($h^{l,{t^\text{(n)}}}_{\boldsymbol \theta}$) to return identity. The superscript ${t^\text{(m)}}$ runs over all the constructed messages, which may include different choices of $w^{l,{t^\text{(e)}}}_{\boldsymbol \theta}$ and $h^{l,{t^\text{(n)}}}_{\boldsymbol \theta}$.

\subsubsection*{Electron embedding update:}
To obtain updated electron embeddings, messages from various edge types are combined and added to a residual connection.
\textsc{DeepQMC} implements a few protocols for the combination of messages, that can be summarized as follows
\begin{equation}\label{eq:edge_embedding}
\mathbf f_i^{(\text{n}),l+1,\text{el}} = \mathbf f_i^{(\text{n}),l,\text{el}}+~
\begin{cases}
    ~\sum_{t^\text{(m)}} g_{\boldsymbol\theta}^{l,{t^\text{(m)}}}\big(\mathbf m_i^{l,{t^\text{(m)}}}\big)\\
    ~g_{\boldsymbol\theta}^{l}\big(\sum_{t^\text{(m)}} \mathbf m_i^{l,{t^\text{(m)}}}\big)\\
    ~g_{\boldsymbol\theta}^l\big(\bigoplus_{t^\text{(m)}} \mathbf m_i^{l,{t^\text{(m)}}}\big) \, ,
\end{cases}
\end{equation}
where $\bigoplus$ refers to the concatenation of the messages. Additionally to the messages constructed according to equation \eqref{eq:graphconvolution} the message types ${t^\text{(m)}}$ can include a residual connection $\mathbf f_i^{(\text{n}),l}$ such that the trainable self-interaction of FermiNet and DeepErwin can be reproduced.

In the above outlined general GNN framework a wide variety of ansatzes can be obtained. 
Furthermore, the implementation of \textsc{DeepQMC} and its GNN framework focus on facilitating rapid extensions with new ansatz variants either by exploration within the existing hyperparameter space or by extending it with new features.  

\subsection{Orbital construction}
The entries of the generalized Slater determinant in \eqref{eq:mb_orbital}
are obtained as products of many-body orbitals $\phi_k^p$ and envelopes $\varphi_k^p$. The many-body orbitals are functions of the final equivariant electron embeddings
\begin{equation}
    \phi_k^p(\mathbf r_i, \{\mathbf r^\uparrow\},\{\mathbf r^\downarrow\}) = \kappa_{\boldsymbol\theta}(\mathbf f^{(\text{n}),L}_i),
    \label{eq:many_body_orbital}
\end{equation}
where $\kappa_{\boldsymbol\theta}$ is an MLP applied electronwise, projecting the embedding dimension to the required number of orbitals. For the $\varphi^p_k$ envelopes, \textsc{DeepQMC} implements linear combinations of exponentials centered on the nuclei
\begin{equation}
    \varphi_k^p(\mathbf{r}_i) = \textstyle{\sum}_I\textstyle{\sum}_{\beta_I} \omega^p_{k\beta_I}\exp(-||\boldsymbol{\Sigma}^p_{k\beta_I}(\mathbf{r}_i - \mathbf{R}_I)||^\alpha)\, ,
\end{equation}
where $\omega^p_{k\beta_I}$ and $\boldsymbol{\Sigma}^p_{k\beta_I}$ are trainable parameters and $\beta_I$ indexes the basis function centered on atom $I$. 
The hyperparameter $\alpha \in (1,2)$ represents the choice of Slater type orbitals with $\alpha=1$, and Gaussian type orbitals (GTOs) with $\alpha=2$. \textsc{DeepQMC} provides an option to restrict the envelopes to be isotropic ($\boldsymbol{\Sigma}^p_{k\beta_I} := \sigma^p_{k\beta_I}\cdot \mathbf I$). The GTOs can be initialized from the molecular orbital coefficients of reference solutions with standard quantum chemistry basis sets obtained in PySCF~\cite{sun_recent_2020}.

\subsection{Determinant construction}
Because the antisymmetry of the wave function is required with respect to the exchange of same-spin particles only, Slater determinants in VMC are typically considered block diagonal and are factored into a spin-up and a spin-down component
\begin{equation}\label{eq:factorized_determinants}
    \psi_{\boldsymbol\theta}=\textstyle{\sum}_p c_p \det[\mathbf{A}^{\uparrow,p}(\mathbf r)]\det[\mathbf{A}^{\downarrow,p}(\mathbf r)].
\end{equation}
Additionally, \textsc{DeepQMC} implements the full determinant option explored by Lin et al.~\cite{lin_explicitly_2023}, which constructs a single determinant using both spin-up and spin-down electrons
\begin{equation}
    \psi_{\boldsymbol\theta}=\textstyle{\sum}_p c_p \det[\mathbf{A}^{\upharpoonleft \! \downharpoonright,p}(\mathbf r)].
\end{equation}
It's noted that since the many-body orbitals are not equivariant under the exchange of opposite spin electrons, the full determinant ansatz is still not antisymmetric with respect to these permutations.
Instead, a full determinant can practically be understood as an expansion in multiple spin-factorized determinants, e.g. by relying on the generalized Laplace expansion of determinants to expand $\det[\mathbf{A}^{\upharpoonleft \! \downharpoonright,p}(\mathbf r)]$ according to the rows corresponding to spin-up electrons
\begin{equation}
    \det[\mathbf{A}^{\upharpoonleft \! \downharpoonright}(\mathbf{r})] = \textstyle{\sum}_S\varepsilon_S \det[\mathbf{A}^{\uparrow, S}(\mathbf{r})]\det[\mathbf{A}^{\downarrow, \bar S}(\mathbf{r})] \, .
    \label{eq:full_det_exp}
\end{equation}
Here, $S$ runs over all subsets of the orbitals that contain as many elements as the number of spin-up electrons, $\bar S$ stands for the complement subset of $S$, 
$\mathbf{A}^{\uparrow, S}({\bf r})$ 
denotes the submatrix of $\mathbf{A}^{\upharpoonleft \! \downharpoonright}({\bf r})$ 
formed from the orbitals in $S$ and the spin-up electrons, while $\varepsilon_S$ is the sign of the permutation defined by the subset $S$.
For the block diagonal matrices of \eqref{eq:factorized_determinants}, the determinants for all subsets of spin-up orbitals containing off-diagonal elements evaluate to zero and the sum in \eqref{eq:full_det_exp} reduces to a single product of a spin-up and spin-down determinant.
Note that since the many-body orbitals defined in \eqref{eq:many_body_orbital} are not equivariant with respect to exchanges of electrons with opposite spins, the terms on the right hand side of \eqref{eq:full_det_exp} with different $S$s will in general be unrelated. In practice, it is conceivable that due to the concrete form of parametrization of the many-body orbitals, there remains some structure in the set of factorized determinants, that makes the full determinant more effective than using an equivalent number of spin-factorized determinants formed from independent orbitals.

\subsection{Jastrow factor and cusp correction}
The antisymmetry of the wave function is retained when multiplying it with a global correction term symmetric under the exchange of the same spin particles. This symmetric correction, traditionally called a Jastrow factor, is well suited to introduce known asymptotics to the ansatz. \textsc{DeepQMC} implements a learnable Jastrow factor $\text{e}^J$, where $J$ is computed from the permutation invariant sum of many-body electron embeddings
\begin{equation}\label{eq:learnable_jastrow}
    J = \eta_{\boldsymbol\theta}\big(\sum_i \mathbf f^{(\text{n}),L}_i\big) \,,
\end{equation}
with $\eta_{\boldsymbol\theta}$ again being implemented by an MLP.
Furthermore, \textsc{DeepQMC} provides a fixed Jastrow factor that implements the known asymptotic behavior ~\cite{kato_eigenfunctions_1957} when two electrons approach
\begin{equation}\label{eq:electron_cusp}
    \gamma(\mathbf{r}) = \sum_{i<j}-\frac{\alpha c_{ij}}{1 + \alpha|\mathbf{r}_i - \mathbf{r}_j|} \, ,
\end{equation}
where $c_{ij}$ is $\tfrac{1}{4}$ if the electrons $i$ and $j$ are of the same spin and $\frac{1}{2}$ if the electrons possess opposite spins and the hyperparameter alpha scales the width of the correction term.
If cuspless Gaussian envelopes are used, a similar cusp correction can be employed for the nuclei
\begin{equation}\label{eq:nuclear_cusp}
    \gamma(\mathbf{r},\mathbf{R}) = \sum_{i,I}\frac{\alpha Z_I}{1 + \alpha|\mathbf{r}_i - \mathbf{R}_I|} \, ,
\end{equation}
serving as a simple replacement for the more involved technique utilized by Hermann et al.~\cite{hermann_deep_2020}.

\subsection{Log-representation of the wave function}
The output of the (unnormalized) neural network wave functions typically spans many orders of magnitude, potentially resulting in instabilities due to finite floating-point precision. In order to improve numerical stability, \textsc{DeepQMC} represents wave functions in the log-domain
\begin{equation}
    \psi = \big(\text{sign}(\psi), \text{log} (|\psi|)\big)\,.
\end{equation}
We mitigate over- and underflow problems during the computation of the determinant by performing it directly in the log-domain using the appropriate \texttt{slogdet} function provided by JAX. In order to perform the summation over multiple determinants $\phi^p$ we apply the log-sum-exp trick
\begin{equation}
\begin{split}
    \text{log}\Big(\big|\sum_p \phi^p\big|\Big) = ~\text{max}&\{\text{log}(|\phi^p|)\} \\ + \text{log}& \Big (\big|\sum_p \text{sign}(\phi^p) \exp\big(\text{log}(|\phi^p|) \\ &~~~- \text{max}\{\text{log}(|\phi^p|)\}\big)\big|\Big)\,.
\end{split}
\end{equation}
Note that for the variational principle to remain valid, it is sufficient to ensure the antisymmetry of the trial wave function, and its explicit sign is not needed for the evaluation of any of the quantities involved in the optimization \eqref{eq:loss_gradient}.

\section{Training}\label{sec:training}
In this section, some technical aspects of the variational optimization of deep-learning trial wave functions are discussed.
While these ansatzes are trained within the standard VMC framework, the characteristics of their optimization differ markedly from other VMC ansatzes, mainly due to the greatly increased parameter count introduced by neural networks.
On the other hand, it is also distinct from most other deep-learning settings owing to the unusual complexity of the loss function and the self-supervised setting, where training data is generated in parallel to the optimization.

\subsection{Loss function and gradient trick}
As discussed in Section\,\ref{sec:variational}, VMC relies on the variational principle by optimizing the wave function ansatz to minimize the expectation value of the local energies.
From a machine learning perspective, this translates to considering the loss function
\begin{equation}
\label{eq:loss}
\mathcal L(\boldsymbol\theta)
  =\mathbb E_{\mathbf r\sim\lvert\psi_{\boldsymbol \theta}\rvert^2}
  \big[E_\text{loc}[\psi_{\boldsymbol\theta}](\mathbf r)\big] \, .
\end{equation}
Naively computing the gradient of this loss would involve taking derivatives of the local energies $E_\text{loc}[\psi_{\boldsymbol \theta} ]$ with respect to the ansatz parameters $\boldsymbol \theta$.
However, evaluating the local energy already involves second derivatives of the wave function with respect to the electron coordinates due to the Laplacian in \eqref{eq:molecular_hamiltonian}.
Consequently, this naive gradient computation would necessitate taking mixed third derivatives of the ansatz.

To reduce the computational costs and numerical instabilities associated with higher order derivatives, a different unbiased estimator of the loss gradient is utilized,
\begin{equation}
\label{eq:loss_gradient}
\begin{split}
    \boldsymbol\nabla_{\boldsymbol\theta}\mathcal L(\boldsymbol\theta)
  = 2\mathbb E_{\mathbf r\sim\lvert\psi_{\boldsymbol \theta}\rvert^2}\big[&
    \big(E_\text{loc}[\psi_{\boldsymbol\theta}](\mathbf r)\\
    &-\mathcal L(\boldsymbol\theta)\big)
    \boldsymbol\nabla_{\boldsymbol\theta}\log\lvert\psi_{\boldsymbol\theta}\rvert
  \big] \, .
\end{split}
\end{equation}
The derivation of this estimator exploits the hermiticity of the Hamiltonian and is given in full detail by Inui et al.~\cite{inui_determinant-free_2021}.
It replaces the derivatives of the local energy with a simple gradient of the wave function, therefore it is expected to be more efficient and numerically stable to evaluate than the direct gradient computation.

\subsection{Local energy evaluation}
The evaluation of the local energies of the wave function ansatz is by far the most computationally demanding part of the training (and evaluation)
\begin{equation}
\begin{split}
    E_{\text{loc}}[\psi_{\boldsymbol\theta}](\mathbf r) =& -\frac12\sum_i\Big(\frac{\Delta_{\mathbf r_i}\psi_{\boldsymbol\theta}(\mathbf r)}{\psi_{\boldsymbol\theta}(\mathbf r)}\Big) + V(\mathbf r) \\ =& -\frac12\sum_i\Big(\Delta_{\mathbf r_i}\log|\psi_{\boldsymbol\theta}(\mathbf r)| \\&~~~~~~~~~~~~~~~+ \big(\nabla_{\mathbf r_i}\log|\psi_{\boldsymbol\theta}(\mathbf r)|\big)^2\Big) + V(\mathbf r).
\end{split}
\end{equation}
While the potential energy term is very cheap to evaluate, the cost of the Laplacian within the kinetic energy term scales steeply with the number of electrons. In this step, one obtains second derivatives of the wave function with respect to the electron coordinates. We obtain these derivatives of the wave function by applying automatic differentiation in backward-forward mode, which we confirmed to be a good choice in the setting of molecular wave functions. Further discussions regarding the memory bottleneck associated with the Laplacian and details regarding the implementation choices are presented in the Appendix \ref{sec:scaling_appendix}.

\subsection{Pretraining}\label{sec:pretraining}
Choosing initial values for ansatz parameters is a non-trivial question common to many  computational chemistry methods.
One need only think of the sensitivity of the self-consistent iterations to the initial guess in Hartree--Fock (HF) and related methods~\cite{saether_density-based_2017,he_divide_2010,yeager_convergency_1979}.
The case of deep-learning VMC ansatzes is no different -- a random initialization of the neural network parameters according to some of the widely adopted schemes of the machine learning community can lead to the optimization diverging or converging to local minima.
This problem becomes increasingly severe with growing system size, presumably due to the higher-dimensional, more complex wave functions of larger molecules and their intricate nodal structure.

A practical solution to this issue is the initialization of the wave function based on a cheap reference solution.
To that end \textsc{DeepQMC} interfaces with \textsc{PySCF}\cite{sun_recent_2020}, enabling the initialization of wave functions from the coefficients of a preceding HF or multi-configurational self consistent field (MCSCF) calculation. While this allows the direct initialization of the neural network wave function ansatz as introduced by Hermann et al.\cite{hermann_deep_2020}, subsequent work suggested that explicitly incorporating an approximate reference wave function in the model can deteriorate performance\cite{gerard_gold-standard_2022}.
Instead a short, supervised pretraining with respect to a reference solution before the self-supervised variational optimization is recommended. In this step, the many-body orbitals of the ansatz are trained to match the reference by minimizing the pretraining loss
\begin{equation}
\begin{split}
        \mathcal L_p (\boldsymbol\theta) = \mathbb E_{\mathbf r \sim | \psi_{\boldsymbol\theta}|^2}\big[& \sum_{ki}\big(\varphi^\text{ref}_k(\mathbf r_i)\\ &- \phi_k(\mathbf r_i, \{\mathbf r^\uparrow\},\{\mathbf r^\downarrow\})\times \varphi_k(\mathbf r_i)\big)^2 \big] \, ,
\end{split}
\end{equation}
where $\varphi^\text{ref}_k$ are the occupied orbitals of the HF/MCSCF wave function.
Unlike the variational loss of \eqref{eq:loss}, computing $\mathcal L_p$ does not involve evaluating the Laplacian of the ansatz, which means that pretraining requires significantly less computational resources than variational training.
Initialization with pretrained orbitals, as introduced by Pfau et al.\cite{pfau_ab_2020}, improves the convergence properties of the variational training and, if well balanced with the subsequent variational optimization, can even slightly boost the final accuracy, as Gerard et al. recently demonstrated~\cite{gerard_gold-standard_2022}.

\subsection{Gradient clipping}
Despite the utilization of sophisticated gradient estimators and pretraining, the convergence of the variational optimization is still often hindered by outliers in the training batches of local energies.
The existence of these outliers is not surprising, considering that the electrostatic energy is singular when two particles coincide, while the kinetic energy is singular at the nodes of the wave function -- energy contributions that the shape of the wave function precisely levels out in later stages of the training. 
While the outliers are valid contributions to the energy expectation value, their presence can inject a lot of noise into the gradient estimates.
To reduce their contribution to the parameter update, the loss and its gradient (given in \eqref{eq:loss} and \eqref{eq:loss_gradient}) are evaluated using clipped local energies $\hat E^{\mu, \sigma}_\text{loc}$, where $\mu$ is the center and $\sigma$ is the half-width of the clipping window.

Regarding concrete choices for $\mu$ and $\sigma$, some empirical findings have been reported in the related literature.
Investigating transition metal atoms using pseudopotentials, Li and coworkers report~\cite{li_fermionic_2022}, that choosing $\sigma = 10 \times \text{std}(E_\text{loc})$  significantly outperforms all other options they've considered.
More recently, von Glehn et al.~\cite{von_glehn_self-attention_2022} found that centering the clipping window at the median of local energies, and using the mean absolute deviation from the median to determine $\sigma$, improves the training of multiple deep-learning ansatzes.
Considering the practical importance of the clipping mechanism, \textsc{DeepQMC} implements the algorithm of von Glehn et al.~\cite{von_glehn_self-attention_2022}, along with an analogous logarithmically scaling ``soft'' clipping scheme introduced by Hermann et al.\cite{hermann_deep_2020}, but also offers full flexibility to the user in specifying custom clipping functions.

Finally, it should be highlighted that the local energies are only to be clipped for computing the gradient of the loss during optimization.
Since clipping can introduce a bias to the estimate of the energy expectation value, variational energy estimates can only be obtained from unclipped local energies.

\subsection{Optimizer}
\begin{figure}
    \centering
    \includegraphics{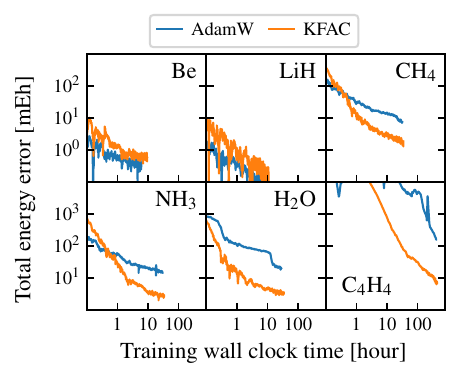}
    \caption{\textbf{Comparing the performance of the AdamW and KFAC optimizers.} Total energy errors during the  training process are shown for beryllium, lithium hydride, methane, ammonia, water, and cyclobutadiene. The horizontal axes show the wall clock time of the training, measured on a single Nvidia GTX 1080 Ti GPU. To obtain smooth training curves, the exponential moving average of the training energy is plotted. While on smaller systems (Be and LiH) AdamW converges slightly faster, due to its lower per-iteration cost, on the larger systems the benefit of using KFAC is clear.}
    \label{fig:optimizers}
\end{figure}

Utilizing natural gradient descent optimization~\cite{amari_neural_1996} or Kronecker-factored approximations thereof~\cite{martens_optimizing_2015} has proven to be a crucial ingredient to the success of variational optimization of deep-learning wave functions on molecular systems~\cite{pfau_ab_2020,spencer_better_2020,gerard_gold-standard_2022,von_glehn_self-attention_2022,pescia_message-passing_2023}.
Consequently, \textsc{DeepQMC} makes use of the Kronecker-Factored Approximate Curvature (KFAC) optimizer implementation of A. Botev and J. Martens~\cite{kfacjax_2022}.
To showcase the importance of the choice of the optimizer, the performance of KFAC is compared to the commonly employed first-order optimizer AdamW~\cite{loshchilov_decoupled_2019}, on variational trainings on six test systems.
The obtained training energy curves are plotted in Figure\,\ref{fig:optimizers}.
To account for the 10--25\% longer per iteration run time of KFAC compared to AdamW, the wall clock time of the training (instead of the usual iteration count) is shown on the horizontal axes. 
The results show that the slightly increased per-iteration cost is offset by the significantly improved per-iteration convergence speed of the KFAC optimizer. 
Furthermore, it is found that the increase in the relative cost of KFAC over AdamW optimization is smaller for systems with larger numbers of electrons. 
In practice, this means that the last percents of correlation energy can be recovered much more efficiently with KFAC, resulting in improved final energies for a given computational budget.

The effectiveness of natural gradient descent in this setting can be rationalized through its connection to the stochastic reconfiguration method~\cite{nomura_restricted_2017, pfau_ab_2020}, known from traditional variational quantum Monte Carlo optimization~\cite{sorella_green_1998, sorella_generalized_2001}.

These higher order methods utilize the Fisher information of the unnormalized density associated with the wave function as a preconditioner to the gradients. 
KFAC extends the application range of natural gradient descent by low-rank approximating the Fisher information, facilitating its computation for neural network wave functions with large numbers of model parameters.

Instead of following the steepest descent in parameter space, an optimization step with the preconditioned gradient is in the direction of steepest descent in distribution space, with distance defined by the Kullback--Leibler (KL) divergence~\cite{martens_new_2020}. 
Considering that in VMC the predicted quantity $\psi_{\boldsymbol \theta}({\bf r})$ directly defines the distribution $p({\bf r}|\boldsymbol\theta) \propto |\psi_{\boldsymbol\theta} ({\bf r})|^2$, one concludes that a natural gradient step is in the direction of maximal loss decrease for a given KL divergence between $\psi_{\boldsymbol\theta}$ and $\psi_{\boldsymbol\theta+\text{d}\boldsymbol\theta}$.
This is an advantageous property, as relying on the KL divergence results in updates that are independent of the way $\psi_{\boldsymbol\theta}$ is parameterized, as opposed to the steepest descent where the Euclidean metric introduces strong dependence.

\section{Sampling}\label{sec:sampling}
An important characteristic of VMC is that the data (electron positions) used to fit the model is generated in tandem with the optimization, by sampling the probability distribution of the electronic degrees of freedom defined by the square of the wave function.
This sampling task comes with its own challenges, due to its tight coupling with the training.
For the variational principle to remain valid, the samples used to evaluate \eqref{eq:expectation_values} must be equilibrated according to the distribution ${\bf r} \sim |\psi_{\boldsymbol\theta}({\bf r})|^2$.
Furthermore, since $\psi_{\boldsymbol\theta}$ is updated in every training iteration, the sampling must account for the corresponding changes in the distribution of the electron positions. 
To carry out this demanding sampling task in a computationally efficient manner, the \textsc{DeepQMC} program package offers two optimized Markov chain Monte Carlo (MCMC) algorithms.
Along with the Random Walk MCMC algorithm~\cite{metropolis_equation_1953,hastings_monte_1970}, referred to as Metropolis sampler, the Metropolis-Adjusted Langevin Algorithm~\cite{besag_discussion_1994} (MALA), referred to as Langevin sampler, is also implemented, that proposes walker updates using overdamped Langevin dynamics.
The implemented MALA includes the correction proposed by Hermann et al.~\cite{hermann_deep_2020}, which scales the electron step sizes around the nuclei to avoid "overshooting" the latter.
Additionally, changes of the wave function during training can be accounted for by re-equilibration after each gradient step or using a batch reweighting scheme.
In the following sections, these MCMC samplers along with the above described sampling difficulties are characterized in more detail.

\subsection{Energy convergence}\label{sec:energy_convergence}
\begin{figure}
    \centering
    \includegraphics{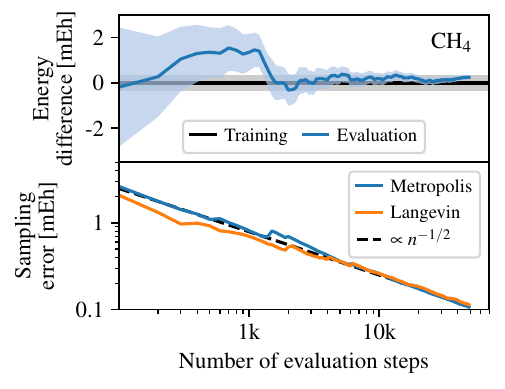}
    \caption{\textbf{Top: Evaluation of a trained wave function ansatz, bottom: sampling error for the evaluation of a trained ansatz.} The convergence of the energy expectation value is depicted during the evaluation of an optimized CH\textsubscript{4} ansatz using two thousand MCMC walkers. In the top pane, sampling is performed with the Random Walk MCMC algorithm. The evaluated energy is compared to the final training energy, with shaded areas showing error estimates. In the bottom pane, the convergence of the sampling errors of evaluations with the Metropolis sampler and the Langevin sampler are compared. The sampling errors converge as $n^{-1/2}$ with the number of samples $n$, as expected from the central limit theorem.}
    \label{fig:sampling_energy}
\end{figure}
First, the convergence of the energy expectation value estimate is investigated, when sampling an unchanging, previously trained ansatz.
In order to draw $n = n_\text{b} \times n_\text{s}$ electron samples $\{\mathbf{r}\}_{ij}$, distributed according to $|\psi_{\boldsymbol\theta}({\bf r})|^2$, a batch of $n_\text{b}$ many walkers is propagated for $n_{\text{s}}$ MCMC steps. Based on the electron samples the energy expectation value is estimated as
\begin{equation}
    \langle E\rangle= \frac{1}{n} \sum_{i=1}^{n_\text{b}}\sum_{j=1}^{n_\text{s}} E_\text{loc}[\psi_{\boldsymbol \theta} ](\{\mathbf{r}\}_{ij})\,.
\end{equation}
Following the central limit theorem\cite{foulkes_quantum_2001, flegal_markov_2008}, the sampling error of such estimates decays proportional to $n^{-1/2}$.
To approximate the sampling error, we utilize the nonoverlapping batch means estimator, as reviewed by Flegal et al.~\cite{flegal_markov_2008}. We first obtain independent estimates of the energy by averaging the local energies over the walker trajectories (batches)
\begin{equation}
\langle E\rangle^i = \frac{1}{n_\text{s}} \sum_{j=1}^{n_\text{s}} E_\text{loc}[\psi_{\boldsymbol \theta} ](\{\mathbf{r}\}_{ij})\,.
\end{equation}
Considering these batches, the sampling error is then estimated as
\begin{equation}
    \langle \sigma_E\rangle  = \sqrt{\frac{\sum_{i=1}^{n_\text{b}} (\langle E\rangle^i - \langle E\rangle)^2}{n_\text{b}(n_\text{b}-1)}} \,.
    \label{eq:sampling_error}
\end{equation}

The convergence of the energy estimate and its error bar throughout the evaluation of an ansatz trained on the CH\textsubscript{4} molecule is plotted in Figure\,\ref{fig:sampling_energy}.
In the top pane, the final value of the exponential walking mean of the training energies, and its estimated error are also shown with a horizontal line and shaded area.
It can be seen from this plot that the energy estimate of the evaluation converges gradually towards the final training energy as expected, while its sampling error converges towards zero. Note that due to the parameter updates during the optimization, the energy estimate from the training is an unreliable estimate and a thorough evaluation of the energy expectation value requires sampling the ansatz with fixed parameters.

On the bottom pane of Figure\,\ref{fig:sampling_energy}, the convergence of the estimated sampling error is compared between the Metropolis sampler and the Langevin sampler.
Importantly, the expected $n^{-1/2}$ convergence behavior is observed for both methods.
Comparing the two algorithms, it can be seen that the error of MALA converges slightly faster than that of Random Walk MCMC, indicating a lower degree of correlation between the subsequent positions of the walkers of the Langevin sampler.

\subsection{Decorrelated sampling}\label{sec:decorrelated_samping}
\begin{figure}
    \centering
    \includegraphics{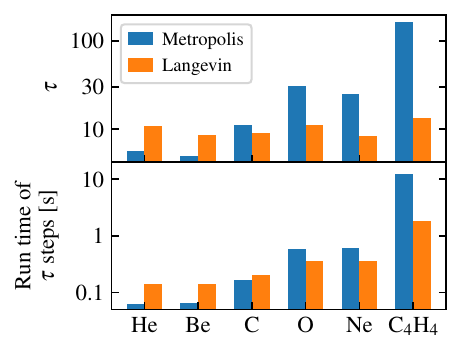}
    \caption{\textbf{Autocorrelation times of the local energy samples.} The top pane shows the MCMC sampling autocorrelation time $\tau$, as defined in Section~\ref{sec:decorrelated_samping}, for a sequence of atoms and the ground state of cyclobutadiene. The bottom pane shows the run time of performing $\tau$ sampling steps for the same systems. The electrons are sampled using either the Metropolis sampler or the Langevin sampler. Following the suggestion of Sokal~\cite{sokal_monte_1996}, the autocorrelation times are estimated using MCMC chains of length $\approx 5\tau$.}
    \label{fig:autocorr}
\end{figure}
To characterize the phenomenon of correlated samples hinted at in Section\,\ref{sec:energy_convergence}, autocorrelation functions of the local energy samples are investigated.
The autocorrelation function of the local energies sampled by a single MCMC chain is defined as:
\begin{equation}
    \rho_{E_\text{loc}}(t) = \int_{-\infty}^\infty \left(E_\text{loc}^{t' + t} - \mu_{E_\text{loc}}\right)\left(E_\text{loc}^{t'} - \mu_{E_\text{loc}}\right)\text{d}t' \, ,
\end{equation}
where $E^{t}_\text{loc}$ denotes the local energy sampled at step $t$, and $\mu_{E_\text{loc}}$ is the mean of the local energies over the entire trajectory. 
The autocorrelation time of the local energy is then computed as $\tau'~=~2 \int_0^\infty \rho_{E_\text{loc}}(t) \text{d}t$. 
Finally, $\tau$ is obtained by taking the mean of $\tau'$s over all propagated MCMC chains, providing a simple measure of local energy autocorrelation.

The autocorrelation times for five atoms of increasing size and the cyclobutadiene molecule are plotted on the upper pane of Figure\,\ref{fig:autocorr}, for both the Metropolis and the Langevin sampler.
The general trend of longer autocorrelation times for larger systems can be observed for the Metropolis sampler.
One of the main causes of this trend is the increasing nuclear charge, which induces higher and higher peaks in the distribution of the electrons near the nuclei.
These pronounced peaks necessitate shorter update proposal radii, ultimately resulting in a higher correlation between subsequent samples.
Furthermore, the autocorrelation time is expected to grow with the increasing complexity of the wave functions and their nodal surfaces.
On the other hand, the Langevin sampler seems less affected by this trend, delivering largely constant autocorrelation times for all systems.
It is reasonable to assume that by explicitly making use of information about the gradient of the wave function, the MALA update proposal retains better decorrelation efficiency than Random Walk MCMC, when considering more and more complicated wave functions.
Finally, the showcased autocorrelation times are in reasonably good agreement with the fact that the default number of decorrelating steps performed between parameter updates is chosen between 10--30 in the currently used neural wave function program packages.

The experiments depicted in Figure\,\ref{fig:autocorr} also demonstrate a slightly smaller correlation between subsequent samples of the Langevin sampler in comparison to those of the Metropolis sampler, for all but the smallest of systems.
On the bottom pane of Figure\,\ref{fig:autocorr}, the wall clock run time of performing $\tau$ sampling steps are shown for each system, to account for the slightly increased computational cost of the MALA update proposal.
Considering wall clock run times, the Metropolis sampler is more efficient on atoms up to carbon, while the Langevin sampler performs slightly to considerably better on the larger atoms and cyclobutadiene. 
While we find MALA to be more efficient than Random Walk Metropolis, we observe that for larger systems with heavier nuclei it could result in less stable optimization.
To improve the black-boxed nature of the method, we applied Random Walk MCMC in all subsequent experiments.

\section{Scaling}
Understanding the scaling of a method's computational cost with the considered system size is of utmost importance in the field of quantum chemistry, where a pervasive caveat of the most accurate approaches is their unfavorable scaling behavior.
Given its high accuracy, the asymptotic scaling of VMC (typically listed with $N^4$)~\cite{foulkes_quantum_2001} is considered favorable.
Although this general scaling is indeed much better than e.g. the $N^7$ scaling of the gold-standard CCSD(T) method, and on par with the scaling of hybrid density functionals (such as DM21~\cite{kirkpatrick_pushing_2021}), deep QMC calculations incur a larger prefactor, resulting in much higher practical costs on systems of intermediate size.
While reducing this prefactor is an important long term goal of method developers in the field, investigating the method's scaling is also of interest, to estimate the prospect of system sizes feasible with further improvement and serve as a baseline for future developments.
In this section, the scaling of the computational cost of the variational training of deep-learning ansatzes is investigated using the \textsc{DeepQMC} program package.
Further scaling aspects of the pseudopotential implementation, and design choices regarding the major computational bottlenecks of the algorithm are discussed in Appendix \ref{sec:scaling_appendix}.

\begin{figure}
    \centering
    \includegraphics{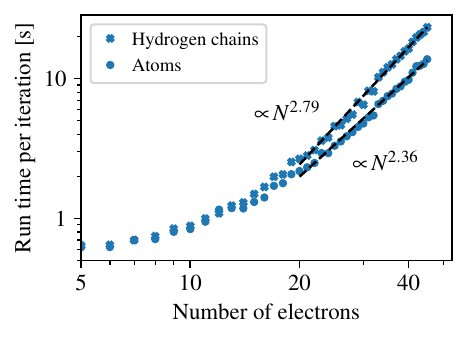}
    \caption{\textbf{Scaling of the computational cost with system size.} The figure depicts the time in seconds per variational optimization step for systems with up to 45 electrons. The timings were obtained for training steps with a batch size of 2000 run, on a single A100 GPU. A power law fit gives the scaling exponent of 2.79 for the hydrogen chains and 2.37 for the atoms respectively.}
    \label{fig:scaling_compute}
\end{figure}
The theoretical scaling of VMC ($N^4$) is obtained when combining the $N^3$ cost of the determinant evaluation with an additional factor of $N$ from the Laplacian required in the computation of the kinetic energy.  
In practice, however, for simulations with the currently feasible system sizes, the determinant evaluation makes up only a fraction of the computational cost, which is instead dominated by the evaluation of the neural networks of the ansatz.
To investigate the practical scaling of a variational training step in \textsc{DeepQMC}, single iteration run times are compared across atoms with increasing atomic numbers, as well as across chains containing an increasing number of hydrogen atoms (Figure\,\ref{fig:scaling_compute}).
Although the systems contain different numbers of particles, due to the parametrization with GNNs the total parameter count of the wave function ansatz changes only marginally between systems. 
On the other hand, owing to the varying numbers of nuclei, isoelectronic species can have slightly different computational requirements.
The system classes of atoms and hydrogen chains were chosen, as they represent the lower and upper bounds respectively, on the number of nuclei a neutral system with a fixed number of electrons can contain.
Consequently, the scaling of the run time with the number of electrons is also expected to be bounded by these system classes.
With the tight empirical bounds of $N^{2.36-2.79}$ depicted in Figure\,\ref{fig:scaling_compute}, the observed scaling of \textsc{DeepQMC} is still far below the theoretical estimate of $N^4$, highlighting the potential for extension to larger systems.

\section{Ansatz validation}\label{sec:ansatz}
Relying on the general framework introduced above, the \textsc{DeepQMC} software suite enables the use of many of the previously published deep-learning ansatzes by providing configuration files to reproduce PauliNet~\cite{hermann_deep_2020}, FermiNet~\cite{pfau_ab_2020} and DeepErwin~\cite{gerard_gold-standard_2022}.
To validate our implementation of these ansatzes, the hyperparameters of sampling, optimization, and GNN architecture are compared in depth to those of the respective reference implementations.
Additionally, it is verified that when using the same parameters, the \textsc{DeepQMC} implementations predict the same wave function value and local energy as their reference counterparts for a given configuration of electrons and nuclei.
Note that we have refrained from exactly matching the cusp-corrected GTOs of PauliNet, because subsequent work has demonstrated that explicitly including a reference solution is limiting the accuracy of the ansatz. However, by combining Gaussian envelopes initialized from the coefficients of a reference calculation with a nuclear cusp correction in the Jastrow factor \eqref{eq:nuclear_cusp} it is possible to obtain a variant of PauliNet within \textsc{DeepQMC}, that matches the characteristics of the original ansatz. 

\begin{figure}[t]
    \centering
    \includegraphics{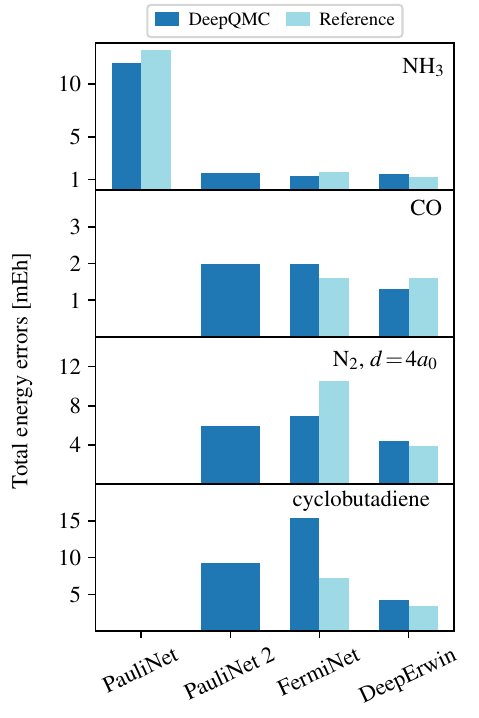}
    \caption{\textbf{Validating the \textsc{DeepQMC} implementation of various ansatzes by comparing their accuracy to published results obtained with their respective reference implementations.} Note that results obtained with the \textsc{DeepQMC} or DeepErwin codebases were computed using 50 000 variational optimization steps and a batch size of 2048, while the FermiNet reference results used 200 000 training steps and 4096 samples in a batch. Results computed with the reference implementations are taken from the works of Pfau~\cite{pfau_ab_2020}, Spencer~\cite{spencer_better_2020}, and Gerard~\cite{gerard_gold-standard_2022}.
}
    \label{fig:ansatz_comparison}
\end{figure}

In Figure \ref{fig:ansatz_comparison} the empirical performance of the various ansatzes is checked against results published in the literature for a small set of molecules. It can be seen that our \textsc{DeepQMC} implementation of PauliNet, FermiNet, and DeepErwin matches the reference energies well.
The remaining discrepancies of FermiNet result from slightly different experimental setups, such as an increased number of reference optimization steps (200 000 compared to 50 000 used here) and batch size (4096 compared to 2048 used here), or an older TensorFlow-based implementation being used in case of N$_2$.
The impact of these changes on the deviations of the model accuracy highlights the importance and difficulty of comparing ansatzes implemented in different codebases under the same experimental conditions.

As a further contribution, we introduce and analyze the performance of a new default ansatz for the \textsc{DeepQMC} program package, which we refer to as PauliNet2.
This exemplary ansatz was optimized to have a good trade-off between accuracy and trainable model parameters. Despite achieving a similar accuracy as FermiNet and DeepErwin for the small systems under investigation (see Figures \ref{fig:ansatz_comparison} and \ref{fig:nemec_corr_ene}), the PauliNet2 ansatz has only about a third of the model parameters of FermiNet and a quarter of DeepErwin (i.e. for the CO molecule 239 829, 766 944, and 998 816 parameters respectively).
The ansatz combines the SchNet-like graph convolutions of PauliNet \eqref{eq:graphconvolution} with the iterative update of the edge embeddings of FermiNet \eqref{eq:edge_embedding}.
Edge features are constructed from difference vectors between the electrons and isotropic exponentials are used as envelopes. Furthermore, the ansatz comprises a trainable Jastrow factor~\eqref{eq:learnable_jastrow} as well as the fixed electronic cusp correction \eqref{eq:electron_cusp}. While these hyperparameters are found to be suitable for the presented experiments, it is conceivable that an extended hyperparameter search targeting specific applications could further improve its performance.
The detailed settings of the discussed ansatzes can be found in the respective configuration files shipped with the \textsc{DeepQMC} package.

\section{Application examples}
In this section, the ease of applying the \textsc{DeepQMC} program package as a black-box method to obtain electronic energies is demonstrated on benchmark datasets.
Two widely different example problems are chosen in order to showcase the general applicability of the presented method.
The experiments are performed using \textsc{DeepQMC} command line interface, which exposes all configuration options of the software suite while also allowing for effortless submission of simple calculations.
Short usage examples of the \textsc{DeepQMC} command line interface are provided in Appendix \ref{sec:usage_appendix}.

\subsection{Small molecule reactions}\label{sec:small_reactions}
The electronic contributions to the reaction energies of 12 reactions involving small inorganic molecules and hydrocarbons are investigated.
These reactions were used by Nemec~\cite{nemec_benchmark_2010} to benchmark the accuracy of Slater--Jastrow (SJ) type trial wave functions, constructed following Drummond et al.~\cite{drummond_jastrow_2004} using electron-electron, electron-nucleus, and electron-electron-nucleus terms in the Jastrow factor.
The 14 participating molecules are built from H, C, O, N, and F atoms, containing at least 2 and at most 22 electrons.
To facilitate the comparison with the DMC results of Nemec~\cite{nemec_benchmark_2010}, the same molecular geometries are considered, obtained from the work of Feller~\cite{feller_survey_2008}.
Reference energies are taken from the review of O'Neill~\cite{oneill_benchmark_2005}.
All electron, complete basis set extrapolated CCSD(T) energies are computed in house, using the PySCF program package~\cite{sun_recent_2020}.

\begin{figure*}[t]
    \centering
    \includegraphics{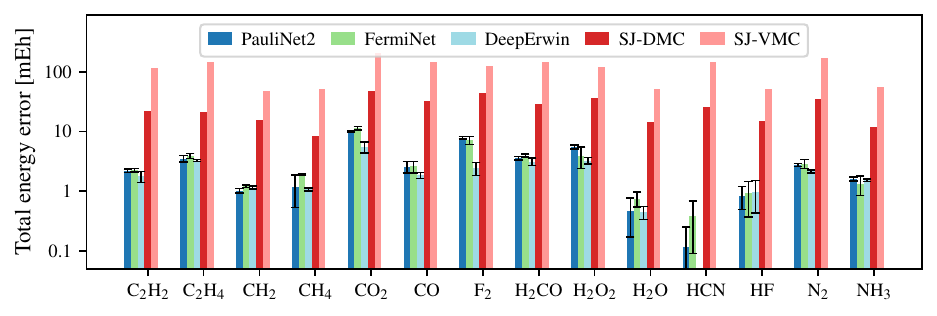}
    \caption{\textbf{Total energy deviations for small molecules of H, C, O, N, and F atoms}. The presented molecules participate in the reactions investigated by Nemec\cite{nemec_benchmark_2010}. Reference energies are taken from the review of O'Neill\cite{oneill_benchmark_2005}, while SJ VMC and SJ DMC results are taken from the work of Nemec\cite{nemec_benchmark_2010}.  Results for the hydrogen molecule are omitted, as it is described nearly exactly by all depicted methods. Error bars denote the sampling error as estimated according to \eqref{eq:sampling_error}.
}
    \label{fig:nemec_corr_ene}
\end{figure*}
First, single-point electronic energies obtained for the participating molecules are compared in Figure\,\ref{fig:nemec_corr_ene}.
On the vertical axis, the error of the recovered total energy is plotted, for VMC and DMC calculations utilizing SJ type trial wave functions, and for VMC with deep-learning ansatzes.
Looking at Figure\,\ref{fig:nemec_corr_ene}, one can observe that the total energy errors of SJ-VMC ansatzes are consistently above 47 mEh (with a mean of 114 mEh), while the associated DMC errors are in the range of 8 - 50 mEh (26 mEh on average).
In comparison, deep-learning ansatzes exhibit at maximum only 11 mEh total energy error, with a mean deviation of 2.6 mEh.
While the main goal of quantum chemistry methods is to accurately model energy differences, rather than recover exact total energies, it is encouraging to see that \textsc{DeepQMC} and deep-learning ansatzes in general are very competitive in this area.

\begin{figure}
    \centering
    \includegraphics{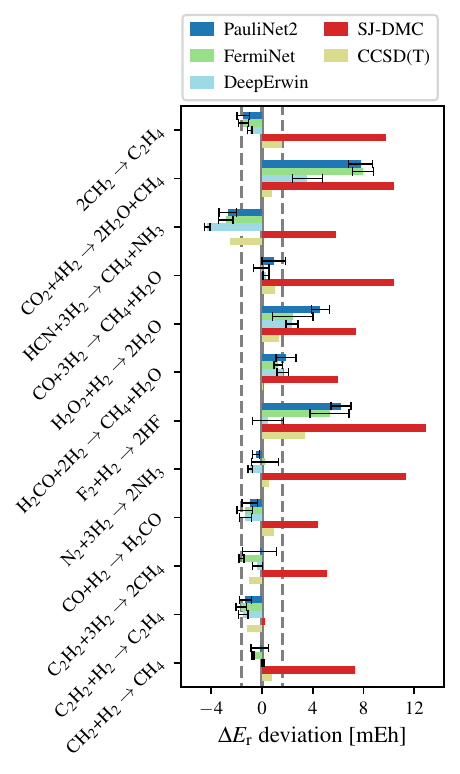}
    \caption{\textbf{Reaction energy deviations for the reactions involving small molecules of H, C, O, N, and F atoms.} Reference reaction energies are computed from the electronic energies reviewed by O'Neill\cite{oneill_benchmark_2005}, SJ-DMC results are obtained from Nemec\cite{nemec_benchmark_2010}, while complete basis set extrapolated CCSD(T) values are calculated in house using PySCF\cite{sun_recent_2020}. The energies for PauliNet2, FermiNet and DeepErwin are obtained with the \textsc{DeepQMC} program package. The range of $\pm 1$ kcal/mol deviation (often referred to as chemical accuracy) is highlighted with dashed lines.}
    \label{fig:nemec_reac_ene}
\end{figure}

The accuracy of the energy differences obtained with SJ-DMC, CCSD(T), and deep-learning QMC methods are compared in Figure\,\ref{fig:nemec_reac_ene}.
Note that energy differences obtained with SJ-VMC are not shown in this figure, as they are an order of magnitude less accurate than the depicted approaches.
Comparing SJ-DMC results with those obtained from \textsc{DeepQMC}, one concludes that combining the VMC method with expressive deep-learning ansatzes greatly increases its accuracy, surpassing SJ-DMC on eleven out of twelve reactions.
The accuracy advantage of \textsc{DeepQMC}'s PauliNet2, FermiNet, and DeepErwin ansatzes is similarly clear when comparing their respective reaction energy mean absolute deviations (MADs) of $2.4$ mEh and $2.3$ mEh, and 1.5 mEh to the 7.6 mEh of SJ-DMC.
As a final comparison, Figure\,\ref{fig:nemec_reac_ene} also shows the reaction energy differences obtained from a complete basis set extrapolated, all-electron CCSD(T).
Not surprisingly, CCSD(T) performs outstandingly on these small, single reference systems in equilibrium geometry, achieving a MAD of 1.3 mEh, and chemical accuracy (less than 1 kcal/mol or 1.6 mEh error) on ten reactions.
In comparison, the MAD value of PauliNet2, FermiNet and DeepErwin for this exemplary study with \textsc{DeepQMC} is found to approach that of CCSD(T), and chemical accuracy is achieved on seven, seven and eight out of twelve reactions respectively.

\subsection{Transition metal oxides}
\label{sec:transition_metal_oxides}
The effects of utilizing pseudopotentials in variational optimization of deep-learning molecular wave function are evaluated on a series of four first-row transition metal oxides.
The bond lengths of the ScO, TiO, VO, and CrO molecules are taken from the experimental results of Annaberdieyev et al.~\cite{annaberdivev_new_2018}.
The latest ccECP pseudopotentials~\cite{annaberdivev_new_2018} are applied to the transition metal atoms only, replacing neon-like cores of 10 electrons.
Although replacing argon cores (18 electrons) with pseudopotentials would result in even larger computational savings, this is avoided as the third shell electrons are known to play a non-negligible role in the bond formation of transition metal atoms~\cite{dolg_energy-adjusted_1987}.
Apart from the introduction of pseudopotentials, the ansatz employed on small molecule reactions (Section\,\ref{sec:small_reactions}) is utilized here without further modifications.

\begin{figure}
    \centering
    \includegraphics{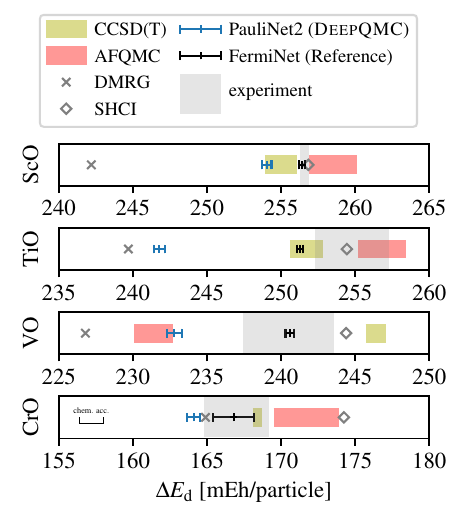}
    \caption{\textbf{Dissociation energy of transition metal oxides calculated using different methods.} \textsc{DeepQMC}+ECP result was obtained using $55\,000$ training steps and $5\,000$ evaluation steps. The results for FermiNet+ECP are taken from Li et al.~\cite{li_fermionic_2022}, where they used 10 times more steps and consequently achieve higher accuracy. Other results are from the references~\cite{dolg_energy-adjusted_1987,williams_direct_2020}.}
    \label{fig:transition_metals}
\end{figure}

Comparing the technical details of pseudopotential calculations to all-electron ones, the advantage of the former is clear.
Due to the exclusion of the fastest moving core electrons, the length of the electron position updates is sixfold increased, and higher accuracy is achieved in a given number of training steps, at about half of the computational cost.
Next, the dissociation energies of the four transition metal oxides are estimated.
The dissociation energy for transition metal X is defined as
\begin{equation}
    \Delta E_\text{d}^\text{XO} = E^\text{X} + E^\text{O} - E^\text{XO}\,,
\end{equation}
where $E^\text{O} = -75.0631(1)\,\text{Ha}$ is the result of an all-electron calculation with the same hyperparameters.
Figure\,\ref{fig:transition_metals} compares the obtained dissociation energies to experimental values~\cite{aoto_how_2017}, and some other accurate computational methods like CCSD(T)~\cite{annaberdivev_new_2018}, FermiNet~\cite{li_fermionic_2022}, auxiliary field quantum Monte Carlo (AFQMC), semi-stochastic heat bath configuration interaction (SHCI), and density matrix renormalization group (DMRG)~\cite{williams_direct_2020}.
Apart from the TiO case, the accuracy of \textsc{DeepQMC} with pseudopotentials is comparable to other theoretical methods, such as CCSD(T) or AFQMC.
The fact that the dissociation energy estimates with \textsc{DeepQMC} are systematically lower than the experimental results, indicates that the single atoms are described more accurately than the oxide molecules. 
This can be counteracted by increasing the expressiveness of the ansatz and investing more compute. 
Note that results obtained with FermiNet~\cite{li_fermionic_2022} utilized a larger ansatz and trained for about ten times more training iterations than done in this study. 

\section{Summary and conclusions}
We have presented the \textsc{DeepQMC} program package -- a general variational quantum Monte Carlo framework for optimizing deep-learning molecular wave functions.
The implementation focuses on modularity, facilitating rapid development of new ansatzes, and provides maximal freedom in choosing the specifics of the variational training setup.
The ansatz shipped with \textsc{DeepQMC} attempts to unify most of the currently existing deep-learning molecular wave functions, while remaining easy to extend as new models emerge.
To reduce the computational complexity associated with heavy nuclei, some popular precomputed pseudopotentials are also implemented.

Using the framework provided by \textsc{DeepQMC}, the most important practical aspects of variational optimization of deep-learning molecular wave functions are discussed.
The importance of a proper gradient estimator along with robust gradient clipping is highlighted.
For consistent ansatz initialization, supervised pretraining to HF wave functions is suggested.
The advantages of using the second-order KFAC optimizer are demonstrated, along with a rationalization of its effectiveness.
The theoretical convergence of the Markov Chain Monte Carlo sampling error is verified, and MALA is shown to be more effective in obtaining decorrelated samples than the widely utilized Random Walk MCMC algorithm.
The empirical scaling of the method's computational cost is found to be more favorable than the most popular post-HF approaches, while its large prefactor is identified as an obstacle on the path to wider adoption.

The black-box application of the program package is demonstrated in two significantly different settings.
The electronic reaction energies of 12 small molecule reactions are computed with a mean absolute deviation of 1.5 mEh, and compared to the 1.3 mEh achieved by CCSD(T) and 7.6 mEh achieved by DMC with SJ ansatzes.
Using the same ansatz hyperparameters, dissociation energies are computed for a series of transition metal monoxides, utilizing the latest ccECP~\cite{annaberdivev_new_2018} pseudopotential.
Improved training characteristics compared to all-electron calculations highlight the benefit of employing pseudopotentials.
The accuracies of the predicted dissociation energies are on par with or exceed those of some other recently popularized methods such as auxiliary field quantum Monte Carlo, or density matrix renormalization group.

To conclude, the presented method shows great promise to become an easy-to-use, general, black-box method accurately describing the molecular electronic structure.
Especially encouraging is the favorable scaling of computational requirements with increasing system size.
It is easy to envision that after further development reducing the large prefactor of the computational costs, the \textsc{DeepQMC} package will prove valuable to the wider quantum chemistry community.

\section*{Data availability}
The data that support the findings of this study will be made openly available upon publication of the manuscript.

\begin{acknowledgments}
We would like to thank Jannes Nys for helpful discussions regarding the nuclear cusp correction, and Leon Gerard for providing reference data for the ansatz comparison.
The work of Z.S. and P.B.S. is funded and supported by the Berlin Mathematics Research Center MATH+ (Projects AA2-8 and AA2-10).
The work of M.M. has received funding from the European Research Council under the European Union’s Horizon 2020 research and innovation program (ERC CoG 772230 "ScaleCell").
Computing time granted by the HLRN and provided on the supercomputer Lise compute at NHR@ZIB is gratefully acknowledged. 

\end{acknowledgments}

\bibliography{deepqmc.bib}

\appendix

\section{Additional scaling experiments}\label{sec:scaling_appendix}

\subsection{Pseudopotentials}
\begin{figure}
    \centering
    \includegraphics{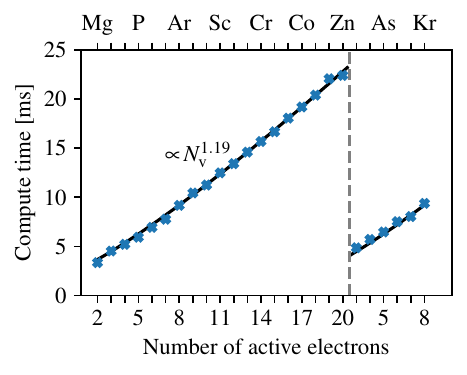}
    \caption{\textbf{Scaling of the compute time of the non-local part of the ccECP pseudopotential.} The evaluation of the second term of Eq. \eqref{eq:pseudopotential} on a single NVIDIA GeForce RTX 3090 GPU. An exponential fit gives a scaling exponent of $1.19 \pm 0.04$ with the number of valence (explicitly treated) electrons. The pseudopotential uses a Neon core for elements up to Zn and a larger core for heavier atoms.}
    \label{fig:scaling_PP}
\end{figure}

Figure \ref{fig:scaling_PP} shows the scaling of the run time of the non-local pseudopotential evaluation (second term in \eqref{eq:pp_integral}) on the third and fourth-row atoms.
This term dominates the total computational overhead of using pseudopotentials overwhelmingly.
From the 5 nested summations of \eqref{eq:pp_integral}, only the sum over the valence electrons scales with the number of electrons, hinting at an approximate linear scaling with system size.
Looking at Figure \ref{fig:scaling_PP}, the obtained empirical scaling of $N^{1.19}$ is in good agreement with expectations.
The sudden jump in run time from 20 to 21 electrons is caused by the reduction of valence electrons, as the utilized ccECP pseudopotentials use a larger core for 4p elements than for 3d ones.

\subsection{Memory requirement}
\begin{figure}
    \centering
    \includegraphics{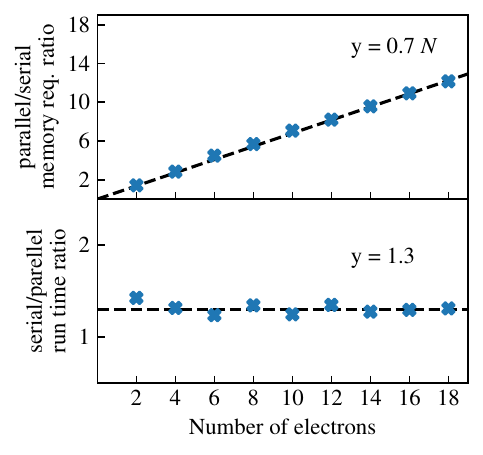}
    \caption{\textbf{Comparing the memory requirements and run times of the serial and parallel Laplacian computations.} Data is obtained by evaluating the Laplacian of an untrained \textsc{DeepQMC} ansatz with respect to all electron coordinates. The parallel implementation computes all diagonal elements of the Hessian at once, while the serial version computes one entry at a time. The number of diagonal Hessian entries scales linearly with the number of electrons. Run times measured on a single Nvidia GTX 1080 Ti GPU.}
    \label{fig:memory_req}
\end{figure}
For all investigated applications, the memory requirement bottleneck is presented by the computation of the Laplacian of the wave function, $\sum^{3N}_i \frac{\partial^2}{\partial^2 r_i} \psi_{\boldsymbol\theta}({\bf r})$.
In this step, one obtains second derivatives of the wave function with respect to the electron coordinates. We obtain the derivatives of the wave function by applying automatic differentiation in backward-forward mode.
While the gradient $\boldsymbol\nabla_{\bf r}\psi_{\boldsymbol\theta}$ is obtained in one backward pass for all coordinates, the diagonal of the Hessian ($\frac{\partial^2 \psi_{\boldsymbol\theta}}{\partial^2 r_i}$) requires an additional $3N$ forward mode differentiations to compute, one for each electron coordinate.
Due to the flexible function transformations of JAX, both the serial and parallel execution of the $3N$ forward mode differentiation passes can be implemented in a few lines of code, with the two implementations presumably differing in how they trade computational efficiency for memory requirement.

To decide between the serial and parallel approach to the Laplacian computation, benchmark calculations on a series of atoms with increasing nuclear charges are performed.
The obtained relative memory requirements of the parallel and serial computations are presented on the left vertical axis of Figure\,\ref{fig:memory_req}.
The observed linear scaling of the relative memory requirement between parallel and serial evaluations can be understood by considering that the parallel implementation holds data for all $3N$ backward passes in memory, while the serial approach stores data for a single pass at a time.
However, the prefactor of the scaling curve is significantly less than three, which indicates that JAX performs some optimizations on the parallel code that reduce the naive $3N$ memory requirement multiplier.
Considering run times of the two versions (lower panel of \ref{fig:memory_req}) it is found that the relative timings of the serial implementation over the parallel implementation do not scale with the system size.
In fact, the ratio of run times between the serial and parallel implementations appears to converge around 1.5.
Taking the above observations into account, the serial implementation of the Laplacian evaluation is chosen, due to its favorable scaling memory requirements which outweigh the slight, non-scaling run time edge of the parallel implementation.

\begin{figure*}[t]
\begin{lstlisting}[language=bash_deepqmc,columns=flexible]
# Optimize the default ansatz for H2O using decorrelated Langevin Sampling and
# a reduced KFAC norm constraint
deepqmc hamil/mol=H2O task/sampler=decorr_langevin task.opt.norm_constraint=0.0005

# Now use a FermiNet ansatz along with its default hyperparameters
deepqmc ansatz=ferminet task=train_ferminet hamil/mol=H2O hydra.run.dir=fn

# Evaluate the previously trained ansatz using 10k inference steps
deepqmc task=evaluate task.restdir=fn +task.steps=10000
\end{lstlisting}
    \caption{\textbf{Example usage of the \textsc{DeepQMC} program package through its command line interface.}}
    \label{fig:cli_usage}
\end{figure*}

\section{Usage of \textsc{DeepQMC}}\label{sec:usage_appendix}
In this section, we provide a few minimal examples of the usage of the \textsc{DeepQMC} command line interface. The interface is based on \textsc{hydra}, which provides a modular way to configure and execute complex jobs. \textsc{DeepQMC} implements a wide variety of configuration options for the wave function ansatz as well as the hyperparameters of training and evaluation. For ease of use, the package includes predefined configuration files, which can be augmented using the command line or extended with custom configuration files. For a thorough tutorial and API documentation the reader is referred to the \textsc{DeepQMC} \href{https://deepqmc.github.io/}{documentation}.
For examples of typical \textsc{DeepQMC} commands see Figure \ref{fig:cli_usage}.

\end{document}